\documentclass[amsmath,11pt]{article}

\usepackage{amsfonts,amsmath,amssymb,graphics,epsfig}

\date{\empty}

\begin{document}

\title{\bf Gravito-electromagnetic resonances\\ in Minkowski space}

\author{Alexandros P. Kouretsis and Christos G. Tsagas\\ {\small Section of Astrophysics, Astronomy and Mechanics, Department of Physics}\\ {\small Aristotle University of Thessaloniki, Thessaloniki 54124, Greece}}

\maketitle

\begin{abstract}
We consider the interaction between gravitational and electromagnetic radiation propagating on a Minkowski background and look into the effects of the former upon the latter. Not surprisingly, the coupling between these two sources leads to gravitationally driven electromagnetic waves. At the second perturbative level, the driving force appears as the superposition of two waves, the properties of which are decided by the initial conditions. We find that the Weyl-Maxwell interaction typically leads to electromagnetic beat-like signals and, in some cases, to the resonant amplification of the driven electromagnetic wave. For physically reasonable initial conditions, we show that these resonances imply a linear (in time) growth for the amplitude of the electromagnetic signal, with the overall amplification also depending on the strength of the driving gravity wave. Finally, we provide order-of-magnitude estimates of the achieved amplification by applying our analysis to astrophysical environments where both gravitational and electromagnetic waves are expected to coexist.\\\\ PACS numbers: 41.20.Jb, 04.30.Tv, 95.85.Sz

\end{abstract}

\section{Introduction}\label{sI}
A typical outcome of wave-wave interactions is a forced vibration, which in some cases may lead to resonances. The latter can in principle facilitate a very efficient energy transfer between the two interacting sources and lead to the substantial amplification of the driven wave. Historically, resonances between gravitational and electromagnetic waves have been studied primarily as a possible means of gravity-wave detection (see~\cite{BM} for a representative though incomplete list). Certain nonlinear aspects of the Weyl-Maxwell coupling were recently investigated by employing covariant techniques, also used here, where the gravitational-wave distortions are monitored through the transverse part of shear perturbations~\cite{T1}. Part of that study assumed a Minkowski background, in which case the gravito-electromagnetic interaction occurs in physical environments where the gravitational field is relatively weak. The analysis, which considered the effects of the Weyl on the Maxwell field and took place at the second perturbative level, showed that gravitational waves could drive and in principle resonantly amplify electromagnetic signals. Also, the nonlinear nature of the coupling, meant that the resonant frequency was determined by those of the originally interacting waves and by their interaction angle.

Resonances occur when the driving and the driven waves oscillate in tune. The most spectacular cases are the so-called undamped resonances, where according to the ``standard picture'' the amplitude of the driven wave diverges. Nevertheless, when physically reasonable/conservative initial conditions are involved, there is no divergence. Instead, the amplitude of the driven wave grows linearly in time (e.g.~see~\cite{LL}). Here, we continue and extend the analysis of~\cite{T1} along these lines. We show, in particular, that the typical outcome of the Weyl-Maxwell resonances identified in~\cite{T1}, is a linear in time growth of the amplitude of the gravitationally driven electromagnetic signal. This means that the longer the interaction lasts, the stronger the amplification. Not surprisingly, the overall effect is also proportional to the strength of the driving gravity wave. An additional result is that the aforementioned resonances are typically preceded by a characteristic phase, during which the amplitude of the driven electromagnetic wave varies periodically. Such periodic variations are known as ``beats'' and in our case occur close to the resonant frequency. More specifically, as the system approaches the point of resonance, there is a series of beat-like signals with progressively increasing duration and amplitude.

The linear (in time) growth of the gravitationally driven electromagnetic wave, at the resonant frequency, makes it easier to estimate its amplification and also the amount of energy transfer from the Weyl to the Maxwell field. The decisive factors are the amplitude of the driving gravitational wave, at the beginning of the interaction, its frequency and the duration of the gravito-electromagnetic resonance. The longer the latter lasts, the stronger the overall effect. Also, the efficiency of the amplification is determined by the product between the initial amplitude and the frequency of the gravitational wave. The larger the amplitude and the higher the frequency, the better. Here, we have applied our results to gravitational radiation with frequency much lower than that of its original electromagnetic counterpart. A condition that is expected to hold in most, if not all, astrophysical situations. The nonlinear interaction terms in the wave equation of the new (the driven) electromagnetic signal allow for the occurrence of resonances even at this case.

Assuming a Minkowski background means confining to environments where gravity is relatively weak. Nevertheless, to maximise the possibility of efficient resonances, we need both gravitational and electromagnetic waves to coexist in relative abundance. The vicinity of compact stellar objects appears to fulfill these requirements. Inspiriling systems of neutron stars and black holes, as well as supernovae explosions and spinning neutron stars, are probably the best candidates, as they are expected to produce high-frequency gravity waves (e.g.~see~\cite{S}). There, it seems feasible to amplify electromagnetic signals by few orders of magnitude, provided that the resonant coupling between the Weyl and the Maxwell field holds for several seconds, or maybe for a few minutes. If so, the mechanism discussed here could provide an additional way of ``extracting'' highly energetic electromagnetic signals from sources like supernovae explosions, for example, which are also expected to produce strong gravitational waves.

\section{The gravito-electromagnetic interaction}\label{sG-EMI}
Consider an empty and static Minkowski spacetime, we allow for the propagation of linear gravitational and electromagnetic radiation. Our aim is to study the interaction between these two sources at the second perturbative level, and in particular how the former affects the latter.

\subsection{Gravitational and electromagnetic waves on Minkowski
space}\label{ssGEMWMS}
Electromagnetic waves are described by the source-free version of Maxwell's equations. On a Minkowski background, these lead to a pair of formalistically identical plane-wave equations for the components of the electromagnetic field. In the framework of the 1+3 covariant formalism, the long-range gravitational field is monitored by the Weyl part of the spacetime curvature. Gravitational radiation, in particular, is described by the transverse components of the electric and the magnetic Weyl tensors ($E_{ab}$ and $H_{ab}$ respectively -- see~\S~1.3.6 in~\cite{TCM} for further discussion and details). However, the high symmetry of the Minkowski space ensures that gravitational-wave perturbations can be described solely by the transverse part of the shear tensor ($\sigma_{ab}$  -- see \S~\ref{ssGIS} below). This gravitationally induced shear also obeys a simple plane-wave propagation equation. All these mean that, to linear order, gravitational and electromagnetic waves propagate according to\footnote{We use geometrised units, with $c=1=8\pi G$, throughout this manuscript.}
\begin{equation}
\sigma_{(k)}= \mathcal{C}\sin(kt+\vartheta) \hspace{10mm} {\rm and} \hspace{10mm} \tilde{E}_{(n)}= \tilde{\mathcal{C}}\sin(nt+\tilde{\vartheta})\,,  \label{linear1}
\end{equation}
respectively (see~\cite{T1} for the details). In the above $k$ and $n$ are the (physical) wavenumbers of the gravitational and the original electromagnetic signal, respectively, while $\vartheta$ and $\tilde{\vartheta}$ are the associated phases. It goes without saying that an expression exactly analogous to (\ref{linear1}b) monitors the linear evolution of the magnetic part of the Maxwell field. We may therefore monitor the propagation of the electromagnetic waves by simply following that of its electric component.

Evaluating expressions (\ref{linear1}a) and (\ref{linear1}b) at $t=t_0=0$, gives $\mathcal{C}=\sigma_0^{(k)}/\sin\vartheta$ and $\tilde{\mathcal{C}}=\tilde{E}_0^{(n)}/\sin\tilde{\vartheta}$ respectively. Then, the linear solutions (\ref{linear1}) recast into
\begin{equation}
\sigma_{(k)}= {\sigma_0^{(k)}\over\sin\vartheta}\, \sin(kt+\vartheta) \hspace{10mm} {\rm and} \hspace{10mm} \tilde{E}_{(n)}= {\tilde{E}_0^{(n)}\over\sin\tilde{\vartheta}}\, \sin(nt+\tilde{\vartheta})\,,  \label{linear2}
\end{equation}
with $\sin\vartheta$, $\sin\tilde{\vartheta}\neq0$ by default. Also, taking the time derivatives of the above, leads to the auxiliary relations
\begin{equation}
\dot{\sigma}_{(k)}= {k\sigma_0^{(k)}\over\sin\vartheta}\, \cos(kt+\vartheta) \hspace{10mm} {\rm and} \hspace{10mm} \dot{\tilde{E}}_{(n)}= {n\tilde{E}_0^{(n)}\over\sin\tilde{\vartheta}}\, \cos(nt+\tilde{\vartheta})\,,  \label{linear3}
\end{equation}
where the overdots indicates differentiation with respect to the observers' proper time. Note that we employ observers living along worldlines tangent to the timelike 4-velocity field $u_a$, with $u_au^a=-1$. Thus, $\dot{\sigma}_{(k)}=u^a\nabla_a\sigma_{(k)}$ and $\dot{\tilde{E}}^{(n)}=u^a\nabla_a\tilde{E}^{(n)}$, where $\nabla_a$ represents the covariant derivative operator, by construction.\footnote{Assuming that $u_a$ is the observers' 4-velocity vector, the symmetric tensor $h_{ab}=g_{ab}+u_au_b$ (with $g_{ab}$ being the spacetime metric) projects into the 3-dimensional space of these observers. Then, ${\rm D}_a=h_a{}^b\nabla_b$ defines the 3-dimensional covariant derivative operator and the symmetric and trace-free tensor $\sigma_{ab}={\rm D}_{(b}u_{a)}-({\rm D}^cu_c/3)h_{ab}$ is the shear associated with the $u_a$-congruence. Also, $E_a=F_{ab}u^b$ defines the electric field vector, as measured in the $u_a$-frame, with $F_{ab}$ representing the antisymmetric electromagnetic (Faraday) tensor.}

\subsection{Gravitationally driven electromagnetic 
waves}\label{ssGDEMWs}
The interaction between gravitational and electromagnetic radiation affects the propagation of both signals. Gravity waves, in particular, can modify the amplitude, the wavelength, the polarisation and the direction of electromagnetic radiation. In the literature, one can find a number of studies addressing this issue from a variety of perspectives (e.g.~\cite{Z}-\cite{To}). Here, we will consider the effects of the Weyl on the Maxwell field and look into the resonant amplification of the latter source by the former. Assuming that the original waves were both monochromatic, the electric component ($E$) of the electromagnetic signal that emerges from the interaction propagates according to the second-order equation
\begin{eqnarray}
\ddot{E}_{(\ell)}+ \ell^2E_{(\ell)}&=& F(n+2k)\sin\left[(n+k)t+\tilde{\vartheta}+\vartheta\right] \nonumber\\ &&-F(n-2k)\sin\left[(n-k)t+\tilde{\vartheta}-\vartheta\right]\,,  \label{nlinear1}
\end{eqnarray}
where $F=\tilde{E}_0^{(n)}\sigma_0^{(k)}/ 2\sin\tilde{\vartheta}\sin\vartheta$. Also, an exactly analogous expression for its magnetic counterpart (see~\cite{T1} for the details). Note that $\ell$, namely the wavenumber of the gravitationally induced electromagnetic radiation, depends on those of the interacting waves and it is given by
\begin{equation}
\ell^2= k^2+ n^2+ 2kn\cos\phi\,,  \label{vphi}
\end{equation}
where $\phi$ represents the interaction angle. According to Eq.~(\ref{nlinear1}), the new electromagnetic signal performs a forced oscillation. Also, the driving agent can be written as the superposition of two plane waves, the characteristics of which depend on those of the originally interacting sources. Not surprisingly, within our adopted perturbative scheme, both the frequency and the phase of the external ``force'' are linear combinations (sums and differences -- namely $n\pm k$ and $\tilde{\vartheta}\pm\vartheta$) of their first-order counterparts.

The outcome of the (second order) gravito-electromagnetic interaction described above follows from the solution of Eq.~(\ref{nlinear1}). The latter can be solved analytically, giving~\cite{T1}
\begin{equation}
E_{(\ell)}= C_1\sin(\ell t)+ C_2\cos(\ell t)+ {F_1\over\ell^2-m_1^2}\,\sin(m_1t+\omega_1)+ {F_2\over\ell^2-m_2^2}\,\sin(m_2t+\omega_2)\,,  \label{nlinear2}
\end{equation}
with $C_{1,2}$ being the integration constants,
\begin{equation}
F_{1,2}= \pm{\tilde{E}_0^{(n)}\sigma_0^{(k)}(n\pm2k)\over2 \sin\tilde{\vartheta}\sin\vartheta}\,, \hspace{10mm} m_{1,2}= n\pm k \hspace{10mm} {\rm and} \hspace{10mm} \omega_{1,2}= \tilde{\vartheta}\pm\vartheta\,.  \label{F1,2}
\end{equation}
The above, together with their magnetic analogues, monitor the evolution of the gravitationally driven electromagnetic radiation, on a Minkowski background, at the second perturbative level. Solution (\ref{nlinear2}) shows that the coupling between the Weyl and the Maxwell fields can lead to the resonant amplification of the latter. This happens when $\ell\rightarrow m_{1,2}$, or equivalently (see expression (\ref{F1,2}b) above) when $\ell\rightarrow n\pm k$. Going back to Eq.~(\ref{vphi}), the latter implies that $\cos\phi=\pm1$ and subsequently $\phi=0,\pi$. In other words, within our framework, resonances take place when the original gravitational and electromagnetic waves propagate in the same, or in the opposite, direction.\footnote{In our analysis we have dropped highly inhomogeneous ``backreaction'' terms from the right-hand side of Eq.~(\ref{nlinear1}) and of its magnetic analogue (see~\cite{T1} for the details). This has allowed for analytical solutions monitoring the gravito-electromagnetic interaction at second order. Keeping the aforementioned terms is unlikely to affect the occurrence of the resonances themselves, but it is likely to modify the setting under which these will occur.}

The possibility of resonances, as a result of the gravito-electromagnetic interaction discussed here, is not surprising. After all, we are dealing with forced oscillations and the latter are known to provide the natural physical stage for resonances to occur. Before turning out full attention to the resonant case, however, it is worth looking at a characteristic  transition phase that takes place as our system approaches the point of resonance.

\subsection{Gravito-electromagnetic beats}\label{ssGEMBs}
The mechanical analogue of the Weyl-Maxwell interaction, as described in Eqs.~(\ref{nlinear1}) and (\ref{nlinear2}), is a harmonic oscillator with two external forces of periodic behavior. Resonances in the system occur within a considerable range of the external frequencies ($m_{1,2}$). This opens the possibility of characteristic signals, which appear close to the resonant frequency and are known as ``beats''. Beats correspond to periodic variations in the amplitude, but at a pace slower than the internal frequency of the system. Following~\cite{LL}, we will demonstrate the occurrence of beats by rewriting solution (\ref{nlinear2}) in the more convenient form
\begin{equation}
E_{(\ell)}= A\cos(\ell t+\varphi)+ B_1\cos\left[{\pi\over2}+(m_1t+\omega_1)\right]+ B_2\cos\left[{\pi\over2}+(m_2t+\omega_2)\right]\,,  \label{bgsol1}
\end{equation}
where now
\begin{equation}
A= \pm\sqrt{C_1^2+C_2^2}\,, \hspace{10mm} B_{1,2}= -{F_{1,2}\over{\ell^2-m_{1,2}^2}}  \hspace{5mm} {\rm and} \hspace{5mm} \tan\varphi= -{C_1\over C_2}\,.  \label{bgsol2}
\end{equation}
Expression (\ref{bgsol1}), which represents a linear superposition of three harmonic oscillations with different amplitudes and frequencies, can be recast into the complex form
\begin{equation}
E_{(\ell)}= \mathcal{A}e^{i\ell t}+ \mathcal{B}_1e^{im_1t}+ \mathcal{B}_2e^{im_2t}\,,  \label{beat1}
\end{equation}
with $\mathcal{A}=Ae^{i\varphi}$ and $\mathcal{B}_{1,2}= B_{1,2}e^{i(\omega_{1,2}+\pi/2)}$. The time variation of the amplitude becomes evident when we rewrite the the last relation as
\begin{equation}
E_{(\ell)}= \left(\mathcal{A}+\mathcal{B}_1e^{i\epsilon_1t} +\mathcal{B}_2e^{i\epsilon_2t}\right)e^{i\ell t}\,,  \label{beat2}
\end{equation}
where $\epsilon_{1,2}=m_{1,2}-\ell$. We also recall that $m_{1,2}=n\pm k$ and $\ell^2=n^2+k^2+2nk\cos\phi$, with $n$ and $k$ representing the wavenumbers of the original electromagnetic and gravitational waves respectively, while $\phi$ is their interaction angle.

\begin{figure}
\begin{center}
\includegraphics[width=0.4\textwidth, angle=0]{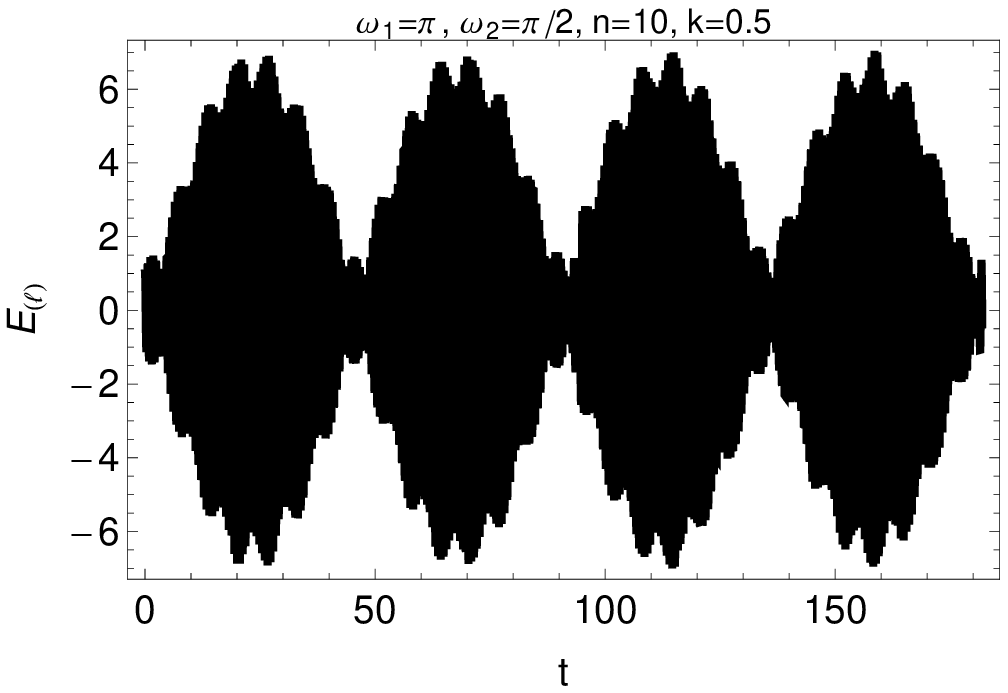}
\includegraphics[width=0.4\textwidth, angle=0]{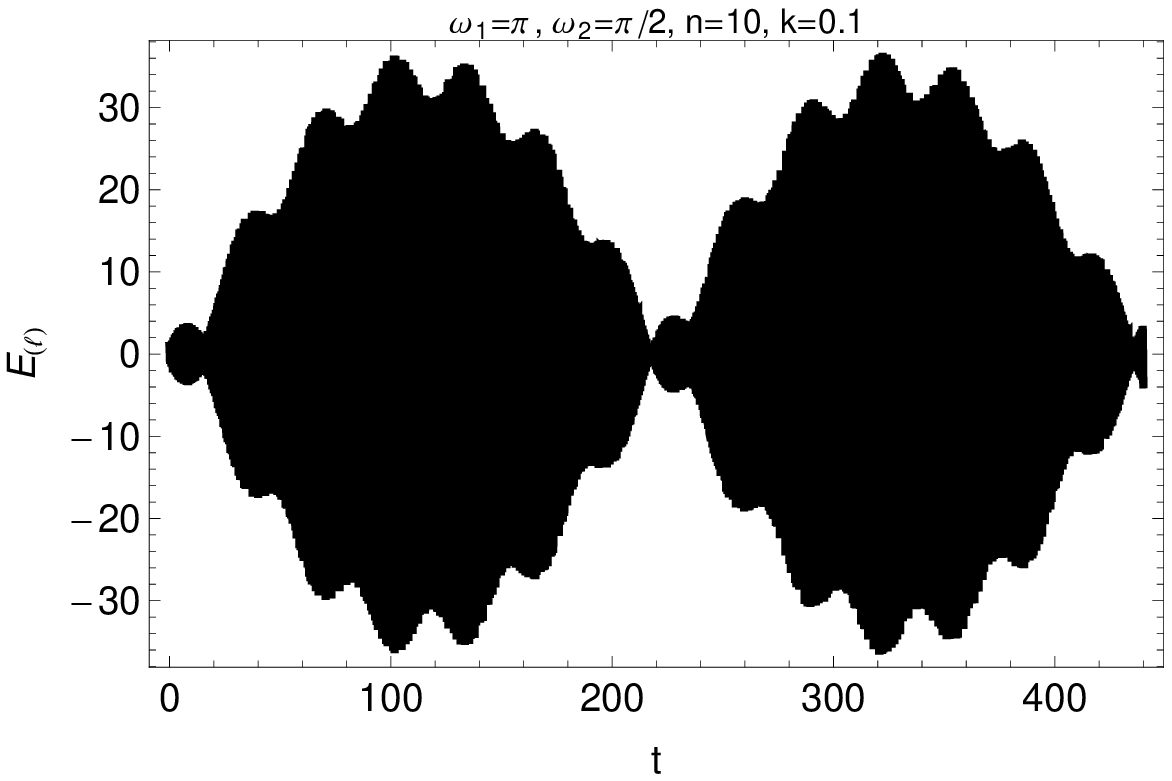}
\caption{\small The gravito-electromagnetic beat, when $n>k$ and $\phi=\pi/4$, as a linear superposition of two simpler beats. Comparing the left to the right panel shows that the duration and the amplitude of the beat increase, both at the same rate, as the difference ($n-k$) between the two wavenumbers grows.}\label{dbeat1}
\end{center}
\end{figure}

\begin{figure}
\begin{center}
\includegraphics[width=0.4\textwidth, angle=0]{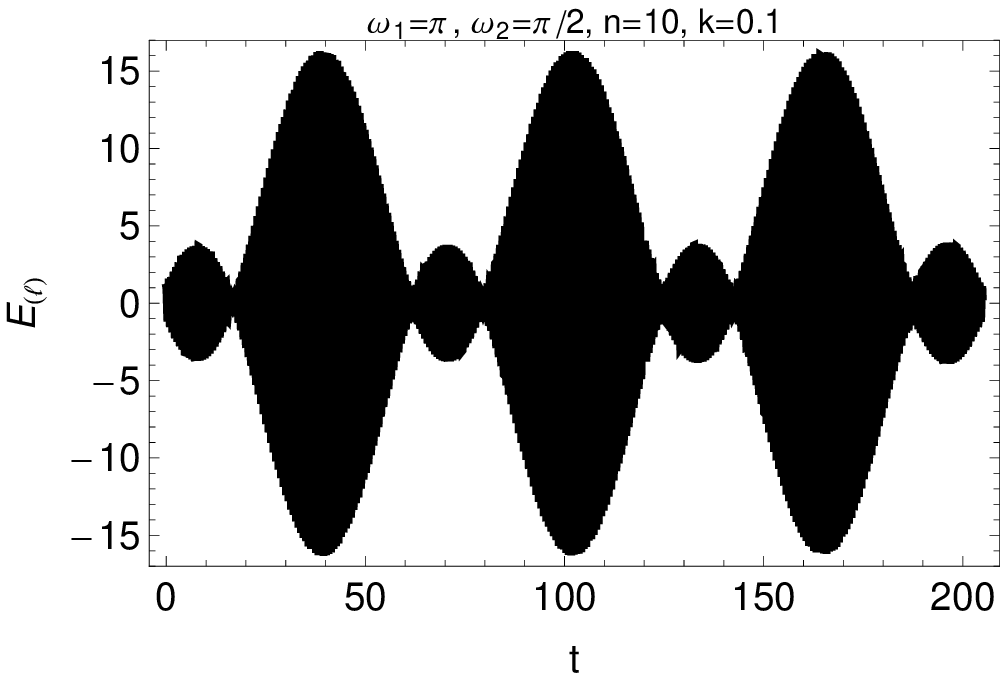}
\includegraphics[width=0.4\textwidth, angle=0]{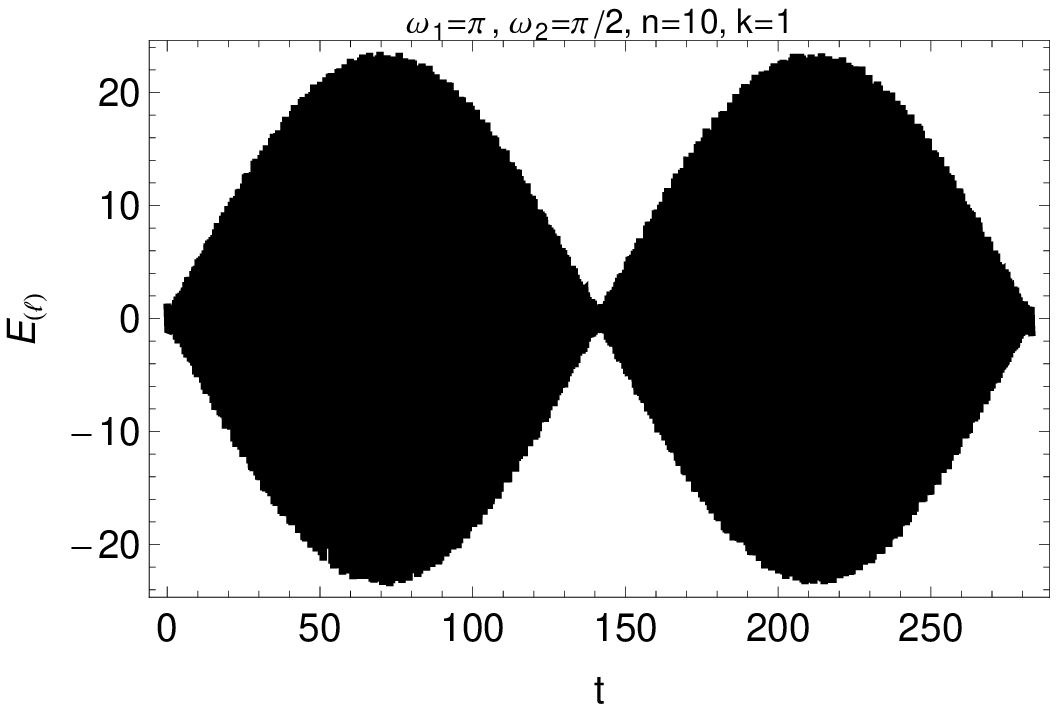}
\caption{\small The two extremal cases for the interaction angle ($\phi$) between the original gravitational and electromagnetic waves (with $n>k$). In the left panel, where the two signals are perpendicular (i.e.~$\phi=\pi/2$), their coupling leads to a double-beat behavior. In the right panel the two wavefronts propagate along the same direction (i.e.~$\phi=0,\pi$) and the driven oscillation appears as a single beat.}\label{dbeat2}
\end{center}
\end{figure}

Let us assume, primarily for the sake of simplicity, that the wavelength of the electromagnetic signal is shorter than that of its gravitational counterpart, namely that $n>k$. Then, keeping up to $k/n$-order terms in the right-hand side of expression (\ref{vphi}), the latter reduces to $\ell=n+k\cos\phi$. This translates into $\epsilon_1= k(1-\cos\phi)$ and $\epsilon_2=-k(1+\cos\phi)$, recasting Eq.~(\ref{beat2}) into
\begin{equation}
E_{(\ell)}= \left[\mathcal{A}+\mathcal{B}_1e^{ik(1-\cos\phi)t} +\mathcal{B}_2e^{-ik(1+\cos\phi)t}\right]e^{i\ell t}\,,  \label{beat3}
\end{equation}
which is the linear superposition of two simpler beats. Changing the interaction angle ($\phi$) between the original two waves, while keeping the rest of the initial features fixed, affects the shape of the beat (see Figs.~\ref{dbeat1} and~\ref{dbeat2}). Modifying the wave frequencies, on the other hand, changes the duration and the amplitude of the beat. The larger the value of wavenumber difference (i.e.~the value of $n-k$), in particular, the longer the duration and the larger the amplitude of the beat. Moreover, the rate of the increase in the duration of the beat matches that in the amplitude (see Fig.~\ref{dbeat1}). Thus, in the extreme case where $n\gg k$ and $n\pm k\rightarrow n$, we should expect a single beat with an amplitude that grows linearly in time (see \S~\ref{ssTnggkC} below).

\section{Gravito-electromagnetic resonances}\label{G-EMRs}
Following (\ref{nlinear2}), when $\ell\rightarrow m_{1,2}$, the amplitude of the gravitationally driven electromagnetic signal diverges. This is probably the most common and best known interpretation of resonances found in many physics textbooks (with some exceptions -- e.g.~see~\cite{LL}). What is less known, is that the outcome of the above described wave-wave interaction also depends on the initial conditions.

\subsection{Setting the initial conditions}\label{ssSICs}
Let us take a closer look at the implications of solution (\ref{nlinear2}). In doing so, it helps to evaluate the integration constants. Taking the time derivative of Eq.~(\ref{nlinear2}), leads to
\begin{equation}
\dot{E}_{(\ell)}= \ell C_1\cos(\ell t)- \ell C_2\sin(\ell t)+ {m_1F_1\over\ell^2-m_1^2}\,\cos(m_1t+\omega_1)+ {m_2F_2\over\ell^2-m_2^2}\,\cos(m_2t+\omega_2)\,.  \label{nlinear3}
\end{equation}
Our next step is to demand that $E_0^{(\ell)}=\tilde{E}_0^{(n)}$ and $\dot{E}_0^{(\ell)}=\dot{\tilde{E}}_0^{(n)}$ at the onset of the gravito-electromagnetic interaction (i.e.~when $t=t_0=0$). Put another way, we assume that the transition from the free electromagnetic wave (prior to the interaction) to the driven one (after the interaction) is smooth.\footnote{Setting $E_0^{(\ell)}=\tilde{E}_0^{(n)}$ is the obvious choice. Otherwise there would have been a discontinuity in the ``position'' of the electric field. It is not necessary to assume that $\dot{E}_0^{(\ell)}=\dot{\tilde{E}}_0^{(n)}$, however, since discontinuities in the ``velocity'' are generally permitted. Allowing for an ``impulse'' at the onset of the interaction between the Weyl and the Maxwell fields, for example, can change the value of $\dot{E}_0^{(\ell)}$ and may thus alter the whole situation considerably.} Under this assumption, expressions (\ref{nlinear2}) and (\ref{nlinear3}) combine to give
\begin{equation}
C_1={\dot{\tilde{E}}_0^{(n)}\over\ell}- {m_1F_1\over\ell(\ell^2-m_1^2)}\,\cos\omega_1- {m_2F_2\over\ell(\ell^2-m_2^2)}\,\cos\omega_2  \label{C1}
\end{equation}
and
\begin{equation}
C_2=\tilde{E}_0^{(n)}- {F_1\over\ell^2-m_1^2}\,\sin\omega_1- {F_2\over\ell^2-m_2^2}\,\sin\omega_2\,.  \label{C2}
\end{equation}
Finally, substituting the above back into Eq.~(\ref{nlinear2}) and then employing some lengthy but fairly straightforward algebra, we arrive at
\begin{eqnarray}
E_{(\ell)}&=& \tilde{E}_0^{(n)}\cos(\ell t)+ {\dot{\tilde{E}}_0^{(n)}\over\ell}\,\sin(\ell t) \nonumber\\ &&+{F_1\cos\omega_1\over\ell} \left[{\ell\sin(m_1t)-m_1\sin(\ell t)\over\ell^2-m_1^2}\right]+ F_1\sin\omega_1 \left[{\cos(m_1t)-\cos(\ell t)\over\ell^2-m_1^2}\right] \nonumber\\ &&+{F_2\cos\omega_2\over\ell} \left[{\ell\sin(m_2t)-m_2\sin(\ell t)\over\ell^2-m_2^2}\right]+ F_2\sin\omega_2 \left[{\cos(m_2t)-\cos(\ell t)\over\ell^2-m_2^2}\right]\,,  \label{nlinear4}
\end{eqnarray}
where $F_{1,2}$, $m_{1,2}$ and $\omega_{1,2}$ are given in (\ref{F1,2}). Clearly, an exactly analogous relation dictates the propagation of the magnetic component of the gravitational induced electromagnetic wave.

Comparing expression (\ref{nlinear4}) to solution (\ref{nlinear2}) in \S~\ref{ssGDEMWs}, we notice that the amplitude of the gravitationally driven electromagnetic wave does not necessarily diverge when $\ell\rightarrow m_{1,2}$. Instead, we are facing an indeterminate situation (of the $0/0$-type), which is the consequence of our adopted initial conditions (see also footnote~4). Following~\cite{LL}, we will bypass the problem of the aforementioned indeterminacy, by appealing to L'Hospital's rule.

\subsection{Linearly growing resonances}\label{ssLGRs}
Of the two possible resonances mentioned in the previous section, let us consider the case where $\ell\rightarrow m_1=k+n$. This occurs when $\phi\rightarrow0$, namely when the original gravitational and electromagnetic waves propagate in the same direction. Then, solution (\ref{nlinear4}) is replaced by
\begin{eqnarray}
E_{(\ell)}\rightarrow E_{(m_1)}&=& \tilde{E}_0^{(n)}\cos(m_1t)+ {\dot{\tilde{E}}_0^{(n)}\over m_1}\,\sin(m_1t) \nonumber\\ &&+ {F_1\cos\omega_1\over m_1} \lim_{\ell\rightarrow m_1}\left[{\ell\sin(m_1t)-m_1\sin(\ell t)\over \ell^2-m_1^2}\right] \nonumber\\ &&+ F_1\sin\omega_1 \lim_{\ell\rightarrow m_1}\left[{\cos(m_1t)-\cos(\ell t)\over \ell^2-m_1^2}\right] \nonumber\\ &&+{F_2\cos\omega_2\over m_1} \left[{m_1\sin(m_2t)-m_2\sin(m_1t)\over m_1^2-m_2^2}\right] \nonumber\\ &&+F_2\sin\omega_2\left[{\cos(m_2t)-\cos(m_1t)\over m_1^2-m_2^2}\right]\,.  \label{resonance1}
\end{eqnarray}
Applying L'Hospital's rule, the above recasts into
\begin{eqnarray}
E_{(\ell)}\rightarrow E_{(m_1)}&=& \left(\tilde{E}_0^{(n)} -{F_1\cos\omega_1\over2m_1}\,t\right)\cos(m_1t)+ \left({\dot{\tilde{E}}_0^{(n)}\over m_1} +{F_1\cos\omega_1\over2m_1^2} +{F_1\sin\omega_1\over2m_1}\,t\right)\sin(m_1t) \nonumber\\ &&+{F_2\cos\omega_2\over m_1} \left[{m_1\sin(m_2t)-m_2\sin(m_1t)\over m_1^2-m_2^2}\right] \nonumber\\ &&+F_2\sin\omega_2 \left[{\cos(m_2t)-\cos(m_1t)\over m_1^2-m_2^2}\right]\,.  \label{resonance2}
\end{eqnarray}
Accordingly, at the $\ell\rightarrow m_1$ limit, the electric (as well as the magnetic) component of the gravitationally induced electromagnetic signal increase linearly in time.

The same linear growth also occurs when $\ell\rightarrow m_2$, which corresponds to $\ell\rightarrow n-k$ and $\phi\rightarrow\pi$ (i.e.~to waves propagating in the opposite direction). Then, after employing L'Hospital's rule once again, Eq.~(\ref{nlinear4}) becomes
\begin{eqnarray}
E_{(\ell)}\rightarrow E_{(m_2)}&=& \left(\tilde{E}_0^{(n)} -{F_2\cos\omega_2\over2m_2}\,t\right)\cos(m_2t)+ \left({\dot{\tilde{E}}_0^{(n)}\over m_2} +{F_2\cos\omega_2\over2m_2^2} +{F_2\sin\omega_2\over2m_2}\,t\right)\sin(m_2t) \nonumber\\ &&+{F_1\cos\omega_1\over m_2} \left[{m_2\sin(m_1t)-m_1\sin(m_2t)\over m_2^2-m_1^2}\right] \nonumber\\ &&+F_1\sin\omega_1 \left[{\cos(m_1t)-\cos(m_2t)\over m_2^2-m_1^2}\right]\,.  \label{resonance3}
\end{eqnarray}

\subsection{The typical $n\gg k$ case}\label{ssTnggkC}
Earlier, in \S~\ref{ssGEMBs}, we considered the $n>k$ case and found that the result of the Meyl-Maxwell coupling was an electromagnetic beat, with a duration and an amplitude that increased (at the same rate) as the wavenumber difference $n-k$ grew larger. Based on these we argued that, at the $n\gg k$ limit, there should be a single beat (instead of a series of beats) with an amplitude that increases linearly in time. Next, we will take a closer look at this claim.

In most realistic situations, in astrophysics for example, the wavelength of the electromagnetic signal is much smaller than that of its gravitational counterpart. This translates into $n\gg k$, in which case we have $m_1=n+k\simeq n$ and $m_2=n-k\simeq n$. At the same limit, relation (\ref{vphi}), implies that $\ell\simeq n$, irrespective of the interaction angle ($\phi$) between the original waves. In other words, when $n\gg k$, the wavenumber of the driving force in the right-hand side of Eq.~(\ref{nlinear1}) and that of the driven electromagnetic signal on the left essentially coincide. Moreover, at the $n\gg k$ limit, expression (\ref{F1,2}a) reduces to
\begin{equation}
F_{1,2}\simeq \pm{\tilde{E}_0^{(n)}\sigma_0^{(k)}n\over 2\sin[(\omega_1+\omega_2)/2] \sin[(\omega_1-\omega_2)/2]}\,, \label{F1,2*}
\end{equation}
since $2\tilde{\vartheta}=\omega_1+\omega_2$ and $2\vartheta=\omega_1-\omega_2$ (see Eq.~(\ref{F1,2}c)).

Taking all of the above into account and substituting (\ref{F1,2*}) into the right-hand side of solution (\ref{resonance2}), we find that, when $n\gg k$, the gravitationally driven electric field is given by
\begin{eqnarray}
E_{(\ell)}\rightarrow E_{(n)}&=& \left(\tilde{E}_0^{(n)}-{F_1\cos\omega_1\over2n}\,t\right)\cos(nt)+ \left({\dot{\tilde{E}}_0^{(n)}\over n}+{F_1\cos\omega_1\over2n^2} +{F_1\sin\omega_1\over2n}\,t\right)\sin(nt) \nonumber\\ &&+{F_2\cos\omega_2\over n} \lim_{m_1\rightarrow m_2} \left[{m_1\sin(m_2t)-m_2\sin(m_1t)\over m_1^2-m_2^2}\right] \nonumber\\ &&+F_2\sin\omega_2\lim_{m_1\rightarrow m_2} \left[{\cos(m_2t)-\cos(m_1t)\over m_1^2-m_2^2}\right]\,,  \label{resonance4}
\end{eqnarray}
with $m_1,m_2\simeq n$. Once again we need to apply L'Hospital's rule. Then, using Eq.~(\ref{F1,2*}) and inserting the initial condition $\dot{\tilde{E}}_0^{(n)}=n\tilde{E}_0^{(n)} \cot[(\omega_1+\omega_2)/2]$ -- see Eq.~(\ref{linear3}b), we arrive at
\begin{eqnarray}
E_{(\ell)}\rightarrow E_{(n)}&=& \tilde{E}_0^{(n)}\left(1+{1\over2}\,\sigma_0^{(k)}t\right)\cos(nt) \nonumber\\ &&-\tilde{E}_0^{(n)} \left[{1\over2}\,{\sigma_0^{(k)}\over n} -\cot\tilde{\vartheta} \left(1+{1\over2}\,\sigma_0^{(k)}t\right)\right]\sin(nt)\,,  \label{resonance5}
\end{eqnarray}
given that $\omega_1+\omega_2=2\tilde{\vartheta}$. The latter is always finite (i.e.~$\sin\tilde{\vartheta}\neq0$ -- see Eq.~(\ref{linear2}b)), which guarantees that  $\cot\tilde{\vartheta}$ does not diverge. Note that the same result can be obtained from Eq.~(\ref{resonance3}) at the $n\gg k$ limit, namely when $m_2\rightarrow m_1\simeq n$. Also, an expression exactly analogous to (\ref{resonance5}) monitors the evolution of the magnetic component of the Maxwell field.

Solution (\ref{resonance5}) describes the resonant growth of the electromagnetic wave that emerges from the interaction between its original counterpart and gravitational radiation of much lower frequency. The result, which holds at the second perturbative level and applies to physical environments where gravity is weak, shows linear (in time) growth for the amplitude of the gravitationally driven electromagnetic signal. Note that, strictly speaking, the  linearity seen in Eq.~(\ref{resonance5}) stems from our initial conditions. These demand a smooth transition from the free to the driven phase of our electromagnetic signal, in accord with the conventional interpretation of wave-wave resonances (see~\cite{LL} and also \S~\ref{ssSICs} here).

\section{Gravito-electromagnetic amplification}\label{GEMA}
According to expression (\ref{resonance5}), the input from the gravitationally induced shear is fixed at the onset of the interaction. This means that the longer the resonance, the more efficient the absorption of gravity-wave energy and the stronger the increase of the electromagnetic signal. In what follows, we will take a closer (and more practical) look at this possibility.

\subsection{Gravitationally induced shear}\label{ssGIS}
Gravitational waves are traveling ripples in the spacetime fabric and their propagation is monitored by the long-range sector of the gravitational field. The latter is described by the electric ($E_{ab}$) and magnetic ($H_{ab}$) components of the Weyl (or conformal curvature) tensor, which obey propagation and constraint equations analogous to Maxwell's formulae (e.g.~see~\cite{TCM}). On a Minkowski background, the transverse part of the aforementioned Weyl tensors are directly related to the transverse component of the shear by means of the linear expressions
\begin{equation}
E_{ab}= -\dot{\sigma}_{ab} \hspace{10mm} {\rm and} \hspace{10mm} H_{ab}= {\rm curl}\sigma_{ab}\,,  \label{Weyl}
\end{equation}
where ${\rm curl}\sigma_{ab}=\varepsilon_{cd(a}{\rm D}^c\sigma^d{}_{b)}$ by definition~\cite{T1}. The former of the above ensures that Weyl curvature distortions induce shear anisotropies. Together, Eqs.~(\ref{Weyl}a) and  (\ref{Weyl}b) imply that we can monitor the propagation of gravitational radiation simply by following that of the shear.

Alternatively, one can relate the shear to the transverse-traceless
part of the metric perturbation ($\mathfrak{h}_{ab}$). More specifically, on a Minkowski background, the perturbed metric is $g_{ab}=\eta_{ab}+\mathfrak{h}_{ab}$ (with $\eta_{ab}={\rm diag}[-1,1,1,1]$) and the distortion in the worldline congruence of observers with 4-velocity $u^a=\delta^a{}_0$ is given by $\nabla_bu_a= \dot{\mathfrak{h}}_{ab}/2$. Then, the transverse-traceless ($TT$) component of the last relation leads to
\begin{equation}
\sigma_{ab}= {1\over2}\,\dot{\mathfrak{h}}^{TT}_{ab}\,,  \label{TTs1}
\end{equation}
which describes the gravitationally induced shear in terms of metric fluctuations. This result translates into the simpler approximate relation
\begin{equation}
\sigma\sim \mathfrak{h}\nu\,,  \label{TTs2}
\end{equation}
with $\mathfrak{h}$ and $\nu$ representing the amplitude and the frequency of the perturbation respectively. As expected, the higher the amplitude and the frequency of the gravitational wave, the stronger the associated shear distortion. Next, we will use the above expression to estimate the gravitationally induced shear and the resulting electromagnetic amplification.

\subsection{Estimating the amplification}\label{ssEA}
The energy density ($\rho$) stored in electromagnetic radiation is proportional to the square amplitude of the Maxwell field. In other words, we may write $\rho\sim E^2$. According to solutions (\ref{linear1}) and (\ref{resonance5}), the amplitudes of the free and the gravitationally driven electric fields ($\tilde{E}_0$ and $E$ respectively) are related by $E\sim\tilde{E} (1+\sigma_0t/2)$, where $\sigma_0$ is the gravito-shear at the beginning of the Weyl-Maxwell coupling. Then, in line with Eq.~(\ref{TTs2}), the overall increase in the energy density of electromagnetic wave is measured by the dimensionless ratio
\begin{equation}
{\rho\over\tilde{\rho}_0}\sim \left(1+ {1\over2}\,\sigma_0t\right)^2\sim \left(1+ {1\over2}\,\mathfrak{h}_0\nu_0t\right)^2\,,  \label{rhoEM1}
\end{equation}
where the zero suffix indicates the onset of the gravito-electromagnetic interaction. Consequently, when $\sigma_0t\gg1$ -- or equivalently for $\mathfrak{h}_0\nu_0t\gg1$, there can be a significant energy transfer from gravitational to electromagnetic radiation. Also, the larger the amplitude and the higher the frequency of the driving gravitational wave, the stronger the amplification effect. Note that the amplitude and the frequency dependencies are decided at the beginning of the gravito-electromagnetic coupling. This means that, within the framework of our analysis, the overall amplification is unaffected by any decay in the strength of the gravitational wave, which may occur during the interaction. Therefore, the longer the resonant coupling between the Weyl and the Maxwell fields lasts, the stronger the residual electromagnetic signal.

In order to operate efficiently, the mechanism described so far ``prefers'' environments were, although gravity is weak, there is a relative abundance of gravitational waves with appreciable strength. The vicinity of compact stellar objects probably provides the most promising possibility. Given that the amplification effect seen in Eq.~(\ref{rhoEM1}) is increasing with frequency, the best astrophysical candidates are probably inspiraling systems of neutron stars and black holes, as well as supernovae or spinning neutron stars. There, the expected gravity-wave frequencies are relatively high, varying between $\sim10$~Hz and $\sim1$~kHz (e.g.~see~\cite{S}). Consequently, for gravitational radiation with amplitude $\sim10^{-3}$ at the beginning of the interaction, the amplification factor ($\mathfrak{h}_0\nu_0$) in the right-hand side of expression (\ref{rhoEM1}) is of order unity.\footnote{Following~\cite{Sc}, we can obtain an order-of-magnitude estimate of the gravitational-wave amplitude. Assuming a virialised system of mass $M$ and size $R$, the amplitude of the emitted wave at a distance $r$ is $\mathfrak{h}\sim M^2/Rr$. Then, setting $R=\alpha R_s$ and $r=\beta R_s$, where $R_s$ is the effective Schwarzschild radius of the source, gives $\mathfrak{h}\sim1/\alpha\beta$. Consequently, for $\alpha\sim10$ and $\beta\sim10^2$, we obtain $\mathfrak{h}\sim 10^{-3}$.} Then, provided the resonant coupling between the Weyl and the Maxwell field holds for a few seconds, the energy density of the electromagnetic signal will grow by roughly two orders of magnitude. If the interaction lasts longer, say for a few minutes, the corresponding increase should range between three and four orders of magnitude.

The theoretical analysis of the previous sections and the rough numerical estimates obtained here, open the possibility of an efficient energy transfer from the gravitational to the electromagnetic sector. Additional insight requires more detailed modeling of the Weyl-Maxwell coupling around strongly gravitating sources, such as those mentioned above. This will undoubtedly  increase the complexity of the problem and thus make it necessary to abandon the strictly analytical treatment and engage numerical methods as well.

\section{Discussion}\label{sD}
Over the years, there has been considerable work on the interaction between gravitational and electromagnetic waves and on the possibility of an energy transfer between them. Here, we have considered the effects of the Weyl on the Maxwell field on a Minkowski background, which corresponds to physical environments where gravity is weak. The aim was to look for analytical solutions that describe gravito-electromagnetic resonances and then investigate their potential implications. In doing so, we used the 1+3 covariant approach to general relativity, within the second-order approximation. Our starting point was two linear plane waves, the electromagnetic and the gravitational, propagating freely on a Minkowski background. Assuming that both waves are monochromatic, we allowed them to interact at the second perturbative level. This induced a new electromagnetic signal, the wave-equation of which corresponds to a mechanical system of a driving oscillator with two external forces (determined by the originally interacting waves). Finding the solutions is a well defined problem of initial conditions and the most characteristic phenomena are \textit{beats} and \textit{resonances}. Here, we found both and provided the corresponding analytical solutions (together with their frequency range) in closed-form expressions. These could potentially lead to detectable gravity-wave imprints in the electromagnetic spectrum.

Initially, the gravito-electromagnetic resonances appear in their familiar (textbook) form, namely as driven oscillations with diverging amplitude. Nevertheless, under physically reasonable/conservative initial conditions, this translates into amplitudes that increase linearly in time. We found that these resonances can occur for a fairly wide range of initial conditions. In fact, the resonant behaviour is recovered even when the gravitational-wave frequency is considerably smaller than that of its original electromagnetic counterpart. This condition is expected to hold for most gravitating sources, since the formation of an event horizon limits the gravity-wave frequency up to $\sim10^4$ Hz. The latter is much smaller than the lowest radio waves observed from various astrophysical sources. The characteristic timescale of the aforementioned linear increase of the electromagnetic wave is determined by the product of the amplitude and the frequency of the driving gravitational radiation. For example, in the favourable scenario where the gravitational radiation is in the kHz band and the interaction occurs within a few Schwarzschild radius from the gravitating source, the energy density of the electromagnetic radiation increases by two orders of magnitude within a few seconds. Also, the linear-growth phase is preceded by a characteristic series of beat-like electromagnetic signals. It is therefore conceivable that, when combined, these effects could lead to an indirect observation of gravitational waves and/or to an electromagnetic follow up in a future gravity-wave detection incident.

The typical candidates for gravito-electromagnetic resonances are astrophysical systems in advanced stages of gravitational collapse. Neutron star and black hole binaries, supernovae and spinning neutron stars are probably the most promising, as they produce high-frequency gravity waves. Also, the predicted energy flux of the gravitational sector is several orders of magnitude above the electromagnetic one. For instance, just before merging, binary systems of several solar masses emit close to $10^{54}$~erg/sec in the form of gravitational radiation. The maximum electromagnetic luminosity during a supernova explosion, on the other hand, is roughly $10^{48}$~erg/sec. We therefore expect the gravitational radiation to dominate and anticipate an abundance of energy in the form of gravity waves. Hence, the interaction discussed here opens the possibility for an efficient transfer of this energy surplus from the gravitational to the electromagnetic sector. Demonstrating that such an energy transfer can readily occur in nature and calculating its magnitude, requires more detailed theoretical models. One would also need to go beyond the analytical limits of this study and involve numerical techniques and codes to deal with the increased complexity of the problem. These will be the next steps of our research effort.\\

\noindent \textbf{Acknowledgments:} The authors wish to thank Nikos Stergioulas for helpful commnets. CGT would also like to thank Axel Brandenburg and Chi-Kwan Chan for several helpful discussions.

\end{document}